\documentclass[journal]{IEEEtran}
\usepackage[utf8]{inputenc}
\usepackage{textcomp}
\usepackage{multirow}
\usepackage{graphicx}

\usepackage{cite}

\ifCLASSINFOpdf
\else
\fi
\begin{document}

\title{Stochastic encoding of graphs in deep learning allows for complex analysis of gender classification in resting-state and task functional brain networks from the UK Biobank}

\author{Matthew Leming,
        John Suckling

\thanks{M. Leming is with the Department
of Psychiatry, University of Cambridge, Cambridge,
Cambridgeshire, CB2 0SZ UK e-mail: ml784@cam.ac.uk.}
\thanks{M. Leming and J. Suckling are with Cambridge University.}}

\maketitle

\begin{abstract}
Classification of whole-brain functional connectivity MRI data with convolutional neural networks (CNNs) has shown promise, but the complexity of these models impedes understanding of which aspects of brain activity contribute to classification. While visualization techniques have been developed to interpret CNNs, bias inherent in the method of encoding abstract input data, as well as the natural variance of deep learning models, detract from the accuracy of these techniques. We introduce a stochastic encoding method in an ensemble of CNNs to classify functional connectomes by gender. We applied our method to resting-state and task data from the UK BioBank, using two visualization techniques to measure the salience of three brain networks involved in task- and resting-states, and their interaction. To regress confounding factors such as head motion, age, and intracranial volume, we introduced a multivariate balancing algorithm to ensure equal distributions of such covariates between classes in our data. We achieved a final AUROC of 0.8459. We found that resting-state data classifies more accurately than task data, with the inner salience network playing the most important role of the three networks overall in classification of resting-state data and connections to the central executive network in task data.\end{abstract}

%
\IEEEpeerreviewmaketitle

\section{Introduction}

\IEEEPARstart{I}{n} recent years, neural networks have proven to be a powerful tool for classification of 2D and 3D images \cite{Krizhevsky2012,Karpathy2014,Maturana2015}. Because of their wide applicability in representing data such as proteins and social networks, much work has been done on adapting neural networks to accept graphs (i.e., networks of nodes interconnected by weighted edges) as input for tasks including whole-graph classification, clustering, and node-wise classification \cite{Bruna2014,Defferrard2016,Hamilton2017,Hechtlinger2017,Kipf2017,Nikolentzos2017}.

Convolutional neural networks (CNNs) adapted for graphs have potentially potent applications in the classification of functional connectivity; a functional MRI reduced to a correlational matrix -- effectively a graph -- that measures the inter-regional relationships between the blood-oxygen-level-dependent (BOLD) signals in predefined anatomical brain areas. While there is no consensus in the neurophysiological interpretation of the resulting networks, certain features have been found to be robust markers of different mental states and disorders; for instance, the default mode network, a large-scale subnetwork within the parietal and frontal areas, has been found to be a marker of resting (task absent) functional connectivity \cite{Raichle2001}.

While other machine learning (ML) models have been developed for analyzing graph data \cite{Jie2013,Kriege2019}, they have often been designed to characterise general networks (such as social networks) rather than fixed-node matrix representations, and so are not ideal for functional connectomes. Additionally, with its utilization of powerful deep learning structures \cite{Kawahara2017,Brown2018}, CNNs are among the most promising ML tools for the diagnosis and prognosis of neurological and mental health disorders using graph representations of the structure and function of the brain. Although they may be applied to classify such graphs, CNNs (and indeed, neural networks more generally) often face a problem with interpretability. Even if CNNs can classify data successfully, it is unknown which features of the input data make a disproportionate contribution in the process, and the model remains a ``black box." Knowledge of such features are especially necessary for biological applications in which the underlying mechanisms of the systems being classified are often of the greatest interest. To overcome the black box problem, a number of ways to visualize and quantify neural networks have been pioneered in recent years. These methods include activation maximization \cite{Erhan2009}, in which the data that maximally activates a hidden node is recorded, occlusion, in which the classification accuracy is measured when specific input data are systematically omitted from the process \cite{Zeiler2013}, and saliency maps \cite{Simonyan2014}, later adapted into class activation maps \cite{Selvaraju2017}, in which the derivative of the neural network with respect to an input datapoint is approximated displaying which parts of the input data effected the most change in the neural network.

The problem of encoding graphs persists in the application of CNNs. Kawahara et al\cite{Kawahara2017} previously employed salience maps in classifying connectivity matrices, using cross-shaped filters in convolutions, to show which connections in the brain had the greatest effect on the resultant classification (thus encoding edge-to-edge connections) instead of square-shaped filters that are more typical for 2D image classification. In our previous work, we used vertical-filters with CNNs and class activation maps to classify functional connectomes \cite{Leming2020}. 

While encoding based on the columns of a connectivity matrix is intuitively sound, given that it accounts for the edges connected to a particular node, it does in theory have three problems. First, the convolutions bias the output class activation maps; a highly salient single edge would also increase the salience of edges in its same row or column. Second, it is difficult to determine the veracity of saliency algorithms from biological data where the ground truth is unknown, and for single runs the algorithms may give spurious results \cite{Kohavi1995}, whereas they often indicate ``visual saliency" for 2D images (i.e., areas of the image on which human subjects focus), which are straightforward to verify by a human observer. Because of the inconsistencies between ML models, the most robust solutions come from averaging salience maps found over a number of trained models \cite{Khosla2018,Leming2020}. Third, convolving whole columns or rows with a single value (node) encodes a large amount of input data that scales with the size of the input matrix. This dilutes the relative contributions of single edges which may be essential in classification, and possibly leads to underfitting.

\subsection{Network brain function across the genders}

Taken on their own, differences found between task-based and resting-state brain activations may be among the most robust discoveries of fMRI studies. The default mode network (DMN) has been consistently identified as a marker of resting-state (i.e. in the absence of a cognitively effortful task) connectomes since it was first described \cite{Raichle2001}. Other brain networks emblematic of particular tasks have been identified as well \cite{Smith2009}, including the dorsal and ventral attention networks \cite{Corbetta2002,Vossel2014}, which are respectively concerned with voluntary focus on features and switches in attention or unexpected stimuli;  i.e., the change between resting-state and task fMRI. As noted by Fox et al\cite{Fox2005}, when performing simple memory tasks, the response commonly observed is proportionally increased activity in certain frontal and parietal cortical regions \cite{Cabeza2000,Corbetta2002} and decreased activity in the posterior cingulate, medial and lateral parietal, and medial prefrontal cortex \cite{Gusnard2001,Simpson2001,Shulman1997,McKiernan2003,Mayozer2001}, which form the default mode network. Fox et al\cite{Fox2005} identified two widely distributed, anticorrelated networks in the brain that exist in the resting state, but intensify during tasks. Additionally, switches between the resting-state and task often involve transitions from the DMN to the central executive (CEN) and salience networks \cite{Goulden2014}. The CEN is the dominant network following suppression of the DMN when a cognitively demanding task is being performed \cite{Fox2006}, while the salience network is activated in a less task-specific manner and more in response to perceived cognitive, homeostatic, or emotional salience \cite{Seeley2007}, which may be brought on by pain, uncertainty, or emotional tasks. Effective connectivity studies with granger causality \cite{Sridharan2008} and dynamic causal modeling \cite{Goulden2014} have indicated that the DMN to CEN transition is modulated by the salience network.

Gender differences in brain networks, and more generally the functional processing of tasks, is an area of active scientific interest. But while functional imaging studies of the brain have often found differences between men and women, it is difficult to compare studies due to small sample sizes, differing analysis methods, different areas selected a priori for testing, and differences in particular tasks. Various task fMRI studies have found widely spread gender differences in the bilateral amygdala, hypothalamus, right cerebellum, and posterior and superior temporal sulcus in response to emotional and visuospatial processing \cite{Hamann2004,Takahashi2006,Mackiewicz2006}; right hemisphere activation in response to visuospatial tests \cite{Gur2000}; differing activations in the superior parietal lobule and the inferior frontal cortex in response to mental rotation tasks \cite{Hugdahl2006}; and limbic regions, prefrontal regions, visual cortex, the anterior cingulate gyrus, and the right subcallosal gyrus in response to emotional faces \cite{Fischer2004,FusarPoli2009}.

Three large sample-size neuroimaging studies that documented functional gender differences in resting-state fMRI in both developing \cite{Tomasi2011,Gur2016} and adult populations \cite{Ritchie2018} found higher local functional connectivity in women than in men, and higher connectivity in the DMN in women and lower connectivity in the sensorimotor cortices, though unlike the emotional stimuli studies there were no particularly localized differences in activation between the samples. This was possibly due to the higher variation of resting-state fMRI due to its unconstrained nature \cite{Buckner2013,Elton2015}. When classifying gender, past ML studies using methods ranging from support vector machines to CNNs, have achieved classification accuracies between 65\% and 87\% \cite{Casanova2012,Satterthwaite2015,Gur2016,Zhang2018a}, depending on the dataset and methods used. In our previous work\cite{Leming2020}, we performed a classification by gender of functional connectomes acquired at multiple sites using a CNN with vertical filters, with a final area under the receiver operating characteristic curve (AUROC) of 0.7680, including an AUROC of 0.8295 with single-site, UK BioBank data. Additionally, DTI data classification has led to exceptionally high accuracies (93\%) \cite{Anderson2019,Xin2019}, though such modalities are not always readily available.

The effects of gender on macro resting-state and task networks are still debated \cite{Goldstone2016}. Some studies \cite{Liu2009,Agcaoglu2015} have found that gender modulates the lateralization of resting-state networks, while other studies have reported only a small \cite{Bluhm2008,LopezLarson2011} or non-significant effect \cite{WeissmanFogel2010,Nielsen2013b}. Network-level gender differences in task fMRI indicate that men and women process tasks differently. Adolescent females have been reported as having higher functional connectivity in the DMN and fronto-parietal networks during a self-referential processing task \cite{Alarcon2018}. Analysis of canonical networks in task fMRI, although not able to draw substantial conclusions on the roles of the networks in different tasks, found that tasks involving fluid intelligence were the most discriminative for gender \cite{Greene2018}. These studies would suggest that men and women process tasks differently. However, they have not been validated on larger datasets.

The objective of this study is to utilize CNNs to classify functional connectomes, but explain the classification performance in terms of those edges and subnetworks that are most salient. To do so, we propose a stochastic deep learning model that allows for the consideration of each edge in a network independently without overfitting, presenting robust results by training and combining many such models. Convolutions with random samples of edges allow for the consideration of each edge independently without overfitting and in training many such models and averaging their outputs we effectively address all of the issues with class activation maps outlined above. Additionally, to overcome the so-called ``black box" problem — that deep learning models are too complex for general interpretation — we use two model visualization methods adapted from deep learning for 2D image analysis to study the role of particular brain networks in the classification. 

\begin{table}[t!]
\caption{Ensemble and Mean AUROCS of all models}
\centering
\begin{tabular}{llllllll}
\multicolumn{2}{l}{\multirow{2}{*}{}} & \multicolumn{2}{c}{All} & \multicolumn{2}{c}{Rest} & \multicolumn{2}{c}{Task} \\
\multicolumn{2}{l}{}                  & Ens.       & Mean       & Ens.        & Mean       & Ens.        & Mean       \\ \hline
\multicolumn{2}{c}{Complete}          & 0.8459     & 0.8010     & 0.8923      & 0.8504     & 0.7683      & 0.7207     \\
\multicolumn{8}{c}{Inner Edges Only}                                                                                  \\
\multirow{2}{*}{CEN}      & Incl.      & 0.8380     & 0.7805     & 0.8844      & 0.8343     & 0.7609      & 0.7027     \\
                          & Excl.     & 0.8386     & 0.7798     & 0.8825      & 0.8315     & 0.7641      & 0.7050     \\
\multirow{2}{*}{DMN}      & Incl.      & 0.8407     & 0.7804     & 0.8868      & 0.8336     & 0.7643      & 0.7018     \\
                          & Excl.     & 0.8420     & 0.7806     & 0.8873      & 0.8334     & 0.7671      & 0.7030     \\
\multirow{2}{*}{SAL}      & Incl.      & 0.8388     & 0.7824     & 0.8860      & 0.8352     & 0.7600      & 0.7050     \\
                          & Excl.     & 0.8392     & 0.7782     & 0.8853      & 0.8308     & 0.7631      & 0.7021     \\
\multicolumn{8}{c}{Connecting Edges}                                                                                  \\
\multirow{2}{*}{CEN}      & Incl.      & 0.8406     & 0.7833     & 0.8872      & 0.8364     & 0.7624      & 0.7059     \\
                          & Excl.     & 0.8287     & 0.7704     & 0.8738      & 0.8228     & 0.7544      & 0.6939     \\
\multirow{2}{*}{DMN}      & Incl.      & 0.8396     & 0.7801     & 0.8836      & 0.8337     & 0.7660      & 0.7020     \\
                          & Excl.     & 0.8278     & 0.7712     & 0.8753      & 0.8246     & 0.7490      & 0.6929     \\
\multirow{2}{*}{SAL}      & Incl.      & 0.8397     & 0.7811     & 0.8875      & 0.8351     & 0.7619      & 0.7024     \\
                          & Excl.     & 0.8321     & 0.7739     & 0.8853      & 0.8253     & 0.7631      & 0.6993     \\

\end{tabular}
\label{tab:table1_ieee}
\end{table}

In this paper, we used CNNs and utilized big data to characterize gender differences in connectomic representations of resting-state and task fMRI (in UK Biobank data, a faces/shapes ``emotion" task \cite{Hariri2002,Barch2013}) with a focus on the DMN, the salience network, and the CEN. We trained our CNNs to classify gender in an extremely large dataset: 16,970 fMRI acquisitions from the UK BioBank, decomposed into multi-wavelet-frequency functional connectivity matrices \cite{Patel2014,Patel2016}. To eliminate the effects of factors such as age, head motion, and intracranial volume, we also detail a multivariate class balancing scheme that ensured equal distributions of these factors within statistical significance. We evaluated performance with the average AUROC, a standard measure of accuracy in ML, across 300 models in an ensemble scheme. We then used guided gradient class activation mapping (Grad-CAM) \cite{Selvaraju2017} and occlusion \cite{Zeiler2013} of individual brain networks to evaluate the salience of each edge within and connecting brain networks, comparing their relative salience within the model.

\section{Methods}
\subsection{Pre-processing}

\subsubsection*{Data acquisition and pre-processing}

The dataset was fMRI data from the UK Biobank, which included both resting-state and task data from a faces/shapes ``emotion" task \cite{Hariri2002,Barch2013}. Details of the acquisition parameters are given elsewhere\cite{Ritchie2018}.

Pre-processing was completed with the fMRI Signal Processing Toolbox (SPT; www.brainwavelet.org). Following initial identification of the brain parenchyma, and affine registration of the 4D sequence to the mean of the sequence, head motion correction was accomplished using SpeedyPP version 2.0. This process utilized AFNI tools and wavelet despiking \cite{Patel2014,Patel2016}, with low- and high-bandpass filters of 0.01Hz and 0.1Hz, respectively, in addition to motion and motion derivative regression. Three motion indicators measured with tools in FSL (FSL motion outliers and FAST; fsl.fmrib.ox.ac.uk/fsl) were recorded that were later applied in class balancing: framewise displacement, spike percentage values, and DVARs. Thus, even if motion correction were imperfect, each dataset would have the same distribution of motion values in either class. 

Time-series at each voxel in the brain were wavelet despiked to remove transient signals, and then functional and structural datasets were registered to Montreal Neurological Institute (MNI) space and parcellated using the 116-area automated anatomical labeling (AAL) template, including subcortical regions \cite{Tzourio2002}, that defined the nodes of the graph.

The average BOLD signal from each parcel was decomposed by wavelet transform in to three frequency bands: 0.05-0.1 Hz, 0.03-0.05 Hz, and 0.01-0.03 Hz. In each frequency band, separately for each dataset, the correlation of the wavelet coefficients between parcels estimated the edge weights resulting in $N(number of datasets)\times 3(wavelet frequency bands)\times 116 (parcels)\times 116 (parcels)$ symmetric connectivity matrices.

Intracranial volume was estimated from structural images with FSL FAST.

Pre-processing was accomplished on a server cluster over a period of several weeks. Due to the volume of datasets, individualized quality control was not possible. From beginning to end, 34.8\% of datasets failed the parcellation/wavelet correlation stages and were rejected from further analysis.

\subsubsection*{Dataset balancing of confounding factors}

When viewed across the full dataset, there were clear differences in the distributions of covariates when stratifying data by both gender and resting-state/task. Gender differences in intracranial volume are well-documented \cite{Ruigrok2014}, and differences in head motion in resting-state and task datasets were also observed. To address these confounding factors, we implemented an algorithm to balance the datasets such that confounding factors, if successfully measured, were not statistically different between groups. This algorithm first required continuous covariates (such as mean framewise displacement, intracranial volume, and age) to be discretised such that values within a given range are placed into ``bins", with each bin covering an equal span of values. Covariates such as collection were already discrete. 

The algorithm curated a subset of the total dataset such that a datapoint from class $A$ within bins ${b_1,b_2,...b_n}$ had a corresponding datapoint within the same multivariate bins from class $B$ that was also within the bins ${b_1,b_2,...b_n}$. In effect, and bearing in mind that males have larger average intracranial volumes, females with smaller intracranial volumes and males with larger intracranial volumes were used less often in the training set, while males with smaller intracranial volumes and females with larger intracranial volumes were more likely to be included in a particular sampling. There is a tradeoff between the size of individual bins and the size of the dataset, since larger bins are naturally more inclusive, but allow for more variation in the distribution of covariates. Thus, the minimum number of bins was used such that it would not reject the null hypothesis with a nonparametric Mann-Whitney U-test with $p > 0.10$. We balanced by age, mean framewise displacement (MFD), intracranial volume (ICV), mean DVARs, and mean spike percentage.

This algorithm was applied twice to our data. The first balanced men and women. This scheme forced a 1:1 ratio between genders, with distributions of respective covariates maintained. Data was then balanced by resting-state and task, though no ratios were forced. This left four divisions in the data: resting-state and task, men and women, with approximately equal distributions of confounding factors.

\subsection{Machine Learning}

We classified functional data by gender. Because classification of UK BioBank rest/task data achieved near-perfect accuracy in our previous work \cite{Leming2020}, we did not repeat this analysis. Here, the focus was on the relative classification accuracy of task data and resting-state data when classifying by gender.

\subsubsection*{Model structure}

The deep learning model was an ensemble of stochastic CNNs. The architecture is shown in Figure \ref{fig:nn_illustration}. We first randomly permuted the columns (nodes) of the connectivity matrices, preserving the permutation order across wavelet frequency bands. These matrices were then input to a CNN with 256 filters of shape $1\times 58\times 1$. This convolved $58\times3$ random values of the matrix which was then fed into three dense layers, each with 64 hidden units, with batch normalization layers, rectified linear unit (ReLU), and 0.5 dropout between them. Finally, the data was binary classified through a softmax layer.

\subsubsection*{Training}

The data were separated into training, validation, and test sets, with an approximate ratio of 4:1:1. We trained 300 CNN models on random class-balanced subsamples of the whole dataset. Each model was trained for 100 epochs (cycles through the training set), and the epoch with the highest validation accuracy was selected. CNN performance was reported on the test set. These 300 models with their respective test set classifications were then unified in an ensemble model. The output classification of a dataset appearing in $\frac{n}{300}$ models was averaged across $n$ models. Thus, datasets were not counted more than once when measuring the final accuracy of the ensemble models, reported as AUROCs. In total, 14,683 datasets were used at least once in the test sets, comprising 86.5\% of the overall dataset.

\subsubsection*{Projection of ensemble upper limit}

The total accuracy of an ensemble model increases with the number of independent models. Assuming an upper limit to the accuracy that can be achieved by adding more models to the ensemble, we measured the AUROC for random samples of 1 to 300 models and fit this relationship to a logarithmic curve ($y = \frac{a}{1 + b e^{-kx}}, k > 0)$, in which $a$ is the upper limit, predicting the accuracy in the limit of a large number of datasets.

\subsection{Visualization of Machine Learning Results}

We used two different ML visualization methods to assess the role of three different, a priori brain networks in the gender classification of resting-state and task data.

\subsubsection*{Brain Network Encoding}

\begin{figure}[!t]
\centering
\includegraphics[width=0.9\columnwidth]{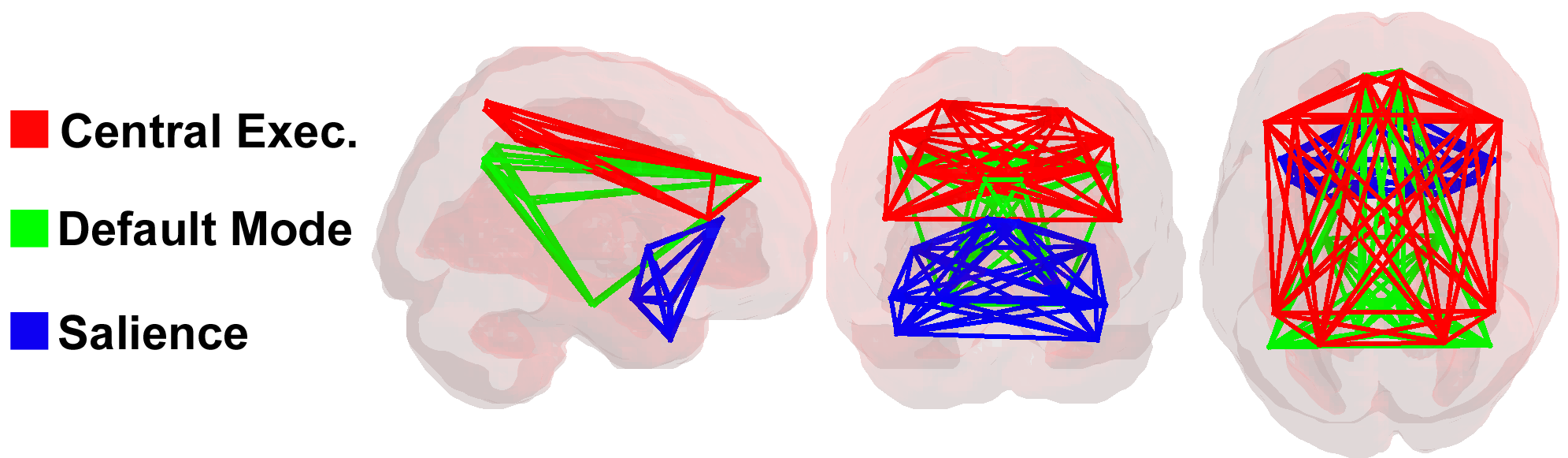}
\caption{A 3d display of the three networks analyzed in this paper, in the AAL parcellation. Green: default mode network; blue: salience network; red: central executive network. Each network is comprised of ten distinct brain regions.}
\label{fig:network_3d_display}
\end{figure}

To assess the role of the DMN, CEN, and salience network in classification, we selected representative nodes from the AAL parcellation (named in Figure \ref{fig:class_activation_maps_histograms}), referring to prior network descriptions\cite{Mulders2015}. Each network was comprised of 10 distinct nodes. The DMN was characterized by a combination of the medial frontal gyrus, posterior cingulum, parahippocampus, precuneus, subgenual anterior cingulate cortex, and inferior parietal lobe, the CEN by the bilateral middle frontal lobe, frontal interior triangularis, frontal superior medial, and the superior and inferior parietal lobe, and the salience network by the bilateral insula, anterior cingulum, amygdala, and the middle and superior temporal pole (Figure \ref{fig:network_3d_display}). 

For both of our analysis methods described below, we isolated edges making up these networks in two different ways: first, by exclusively selecting edges within the network; i.e. edges connecting two nodes of a given network (comprising $\frac{10\times(10-1)}{2}=45$ unique edges); and second, all edges within, and connecting to a network, by selecting those edges that connect to at least one other node (comprising $10\times(116-1) - \frac{10\times(10-1)}{2}=1105$ unique edges). Thus, for each analysis method, two sets of results are presented: one for the sets of edges within a network, and the other for all edges connected to a network.

\subsubsection*{Gradient Class Activation Maps}

We applied the Grad-CAM algorithm \cite{Erhan2009,Selvaraju2017,raghakotkerasvis} to find class activation maps (CAMs) for each dataset in each CNN model. Grad-CAM is an extension of the general salience algorithm\cite{Simonyan2014}. In its simplest form, salience is obtained by taking the derivative (approximated as a first-order Taylor expansion) of a particular deep learning model with respect to a particular input image. In studies of 2D images, CAMs are able to distinguish between different objects within a single image belonging to different classes \cite{Selvaraju2017}; for example, in a multiclass classifier of a picture of a cat and a dog, taking an image with respect to class 0 would highlight the cat, while taking the same image with respect to class 1 would highlight the dog. Grad-CAM extends this by making CAMs applicable to a variety of CNNs, including those that use fully-connected deep layers, as used here.

We derived CAMs from each independent CNN with respect to both class 0 (females) and class 1 (males) across three wavelet bands and averaged these across the 300 models, producing a single $116\times 116$ CAM for each fMRI dataset in the ensemble models. The total distribution for CAM values within and connecting to each particular brain network was then compared to every other CAM value. Due to the extremely large number of values, distributional differences were measured by Cohen's D (effect size), rather than statistical significance.

\subsubsection*{Occlusion}

In separate gender classification models, we occluded half of the edges for each model in the ensemble and trained on the occluded data. This was inspired by photographic image occlusion \cite{Zeiler2013} which deliberately excludes portions of data and measures relative classification accuracy with the occluded data as a means of detecting salient areas. The importance of the three brain networks to the classification was tested by comparing the average AUROC of $300$ models whose occluded edges were the edges making up the particular brain network, and $300$ models for which brain networks were not occluded. We trained on each set using the same 300 model/ensemble scheme detailed above (see Figure \ref{fig:occlusion_histograms}, top). The relative accuracies of these independent models, both on the complete dataset and for the resting-state and task fMRI data, were compared to understand the contributions of different networks to gender classification in both resting-state and task fMRI. In particular, we applied a nonparametric statistical test on the two sets of 300 AUROCs including and excluding a particular brain network, then reported the p-value of this test, corrected for multiple comparisons.

We trained, for each of our three networks, 300 models that included the given network and 300 excluding it, each with the two different encoding schemes (i.e. considering the edges only within a network and all edges connected to a network), for each of the three networks (DMN, CEN, and salience network). In total, we trained $2\times 2\times 3\times 300=3600$ models for these occlusion tests.

\section{Results}

\subsection{Datasets and pre-processing}

\subsubsection*{Dataset Balancing}

\begin{figure}[!t]
\centering
\textbf{Unbalanced}\\
\includegraphics[width=0.45\columnwidth]{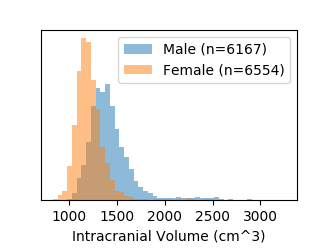}
\includegraphics[width=0.45\columnwidth]{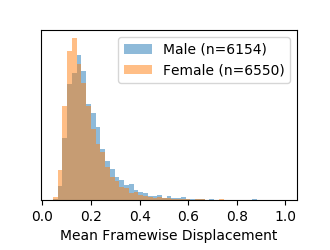}\\
\includegraphics[width=0.45\columnwidth]{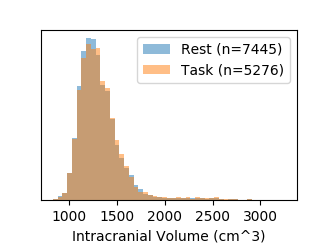}
\includegraphics[width=0.45\columnwidth]{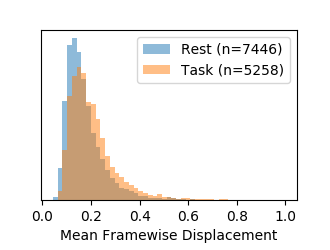}\\
\textbf{Balanced}\\
\includegraphics[width=0.45\columnwidth]{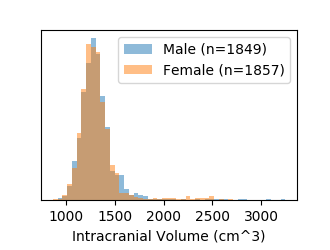}
\includegraphics[width=0.45\columnwidth]{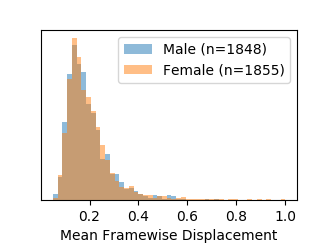}\\
\includegraphics[width=0.45\columnwidth]{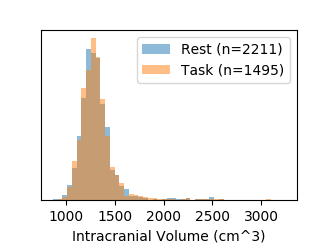}
\includegraphics[width=0.45\columnwidth]{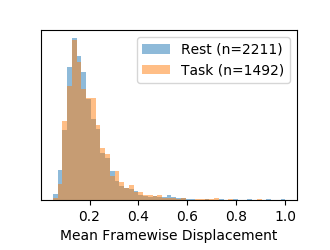}
\caption{Histograms displaying distributions of random training sets with respect to mean FD and intracranial volumes, divided both by gender and resting-state/task, before and after the class balancing scheme.}
\label{fig:mfd_icv_histograms}
\end{figure}

The datasets displayed significant motion effects between groups, especially with regards to task- and resting-state differences, as well as significant differences in intracranial volumes between genders (Figure \ref{fig:mfd_icv_histograms}). The class balancing scheme selectively eliminated datasets such that each class had similar distributions across each covariate, as well as a 1:1 ratio of males to females. The same balancing procedure was also performed for resting-state and task data, with the original ratios present in the dataset maintained. Class balancing disincentivized the model from classifying based on confounding factors. The balanced class distributions can be seen at the bottom of Figure \ref{fig:mfd_icv_histograms}.

\subsection{Machine Learning}

\subsubsection*{Model Accuracy}

\begin{figure}[!t]
\centering
\includegraphics[width=2.5in]{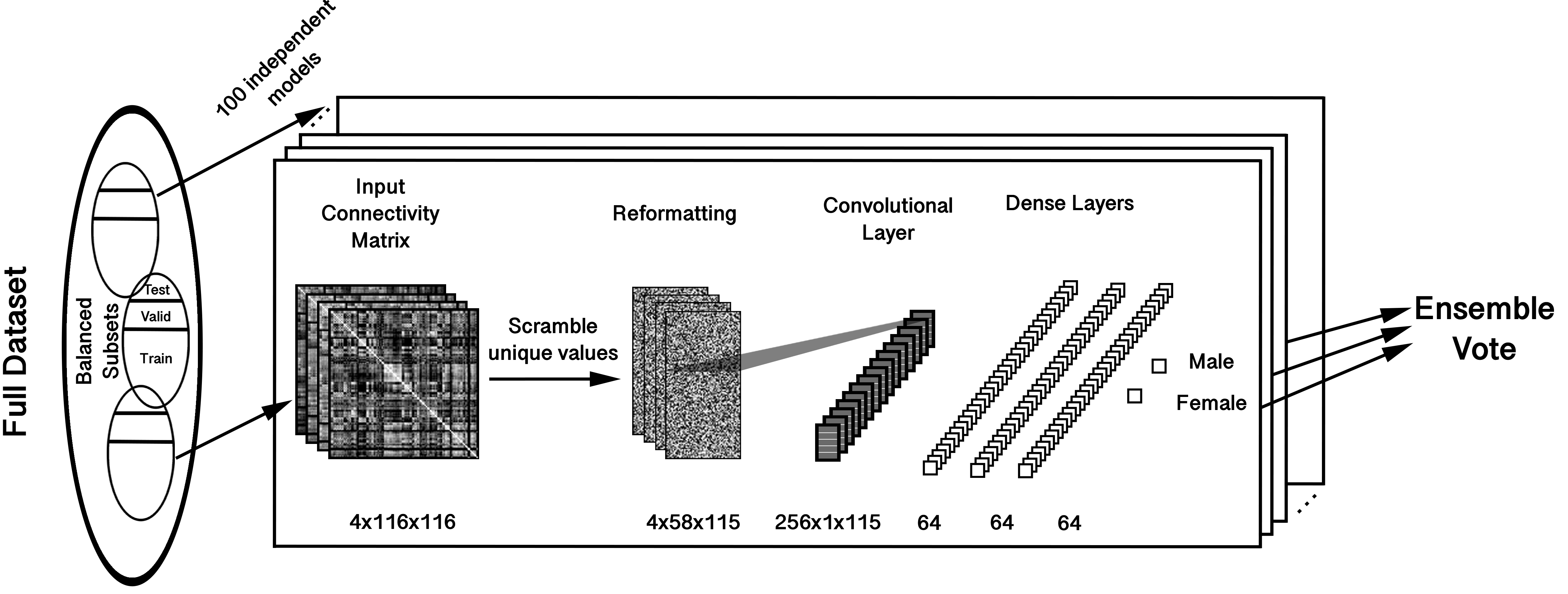}
\caption{In this model, matrices are encoded by random scrambling prior to being fed into a single convolutional layer, followed by three dense layers. In between each layer is a batch normalization and rectified linear unit (ReLU) layer, with 50 percent dropout in between the dense layers. Our training scheme trains 300 such models, with with its unique scrambling order, independently on a class- and covariate-balanced subset of the whole dataset, then combines votes for datapoints appearing in overlapping test sets into a final ensemble vote.}
\label{fig:nn_illustration}
\end{figure}
\begin{figure}[!t]
\centering
\includegraphics[width=0.95\columnwidth]{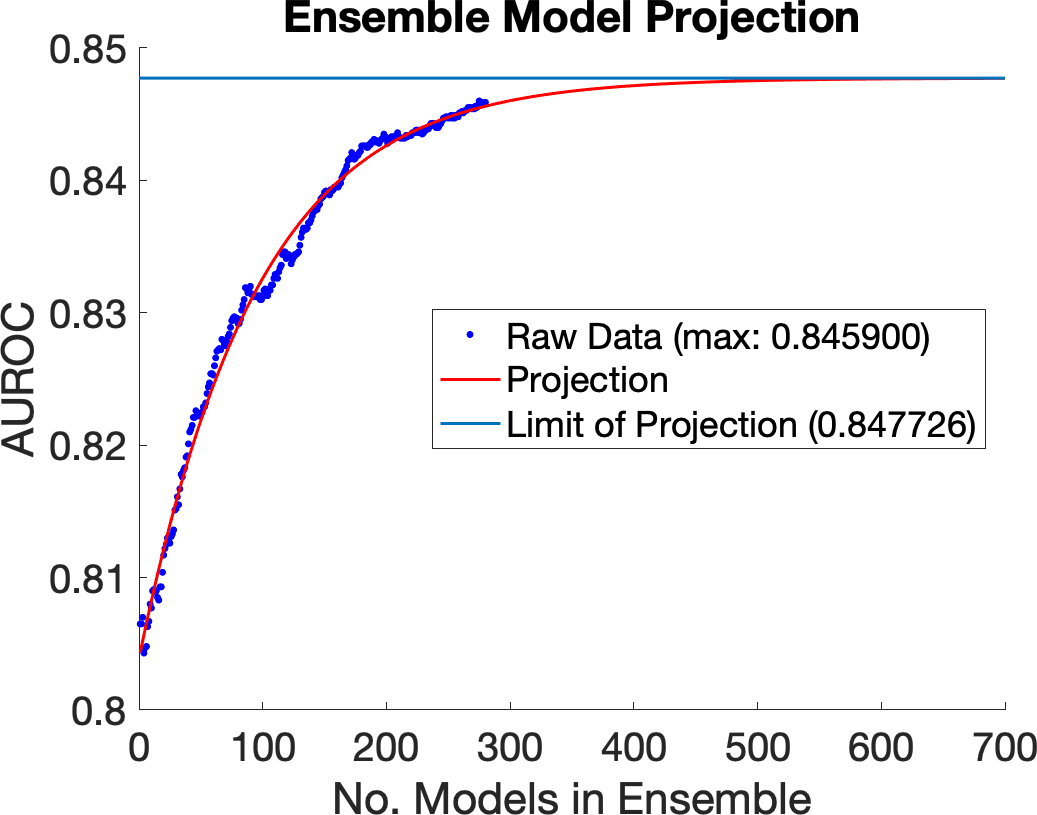}
\caption{Gender classification AUROC across 1 - 300 independent CNNs included in the ensemble model. The raw data is plotted, as well as the projection of this trend using a logistics growth model ($y = \frac{a}{1 + b e^{-kx}}, k > 0)$, which assumes a hard upper limit ($a$) to the classification accuracy that can be achieved by simply increasing the number of models in the ensemble. The model predicts that simply adding more models to the ensemble beyond 300 achieves limited returns. The upper limit is 0.8477, with 95\% confidence bounds between 0.8473 and 0.8481.}
\label{fig:ensemble_model_projection}
\end{figure}

We initially classified by gender balanced datasets with both resting-state and task fMRI. We used 300 independent CNNs that took as input randomly scrambled unique values of the input wavelet correlation matrices (Figure \ref{fig:nn_illustration}) in a stratified cross-validation \cite{Kohavi1995} scheme. The final results for the 300 models are given in Table \ref{tab:table1_ieee} (top row) with an average AUROC of 0.8010 when assessing the CNNs independently. However, when all 300 models were aggregated into a single classification such that predictions for a particular dataset appearing across multiple independent models were averaged into a single value (Figure \ref{fig:nn_illustration}), the AUROC was 0.8459. 

The ensemble model also classified gender in resting-state fMRI with an ensemble AUROC of 0.8923 and task fMRI with an AUROC of 0.7683, a difference of 0.1240. Full results are given in Table \ref{tab:table1_ieee}.

\subsubsection*{Projection of ensemble upper limit}

The upper predicted limit of AUROC in the limit of a large number of datasets, based on a logarithmic model, is shown in Figure \ref{fig:ensemble_model_projection}, and was found to be 0.8477.

\subsection{Visualization of Machine Learning Results}

\subsubsection*{Gradient class activation maps}
\begin{figure*}[!t]
\newcommand\RotText[1]{\rotatebox{90}{\parbox{3cm}{\centering#1}}}
\fontfamily{phv}\fontsize{12}{6}\selectfont
\centering
\hfil
\includegraphics[height=0.20\textwidth]{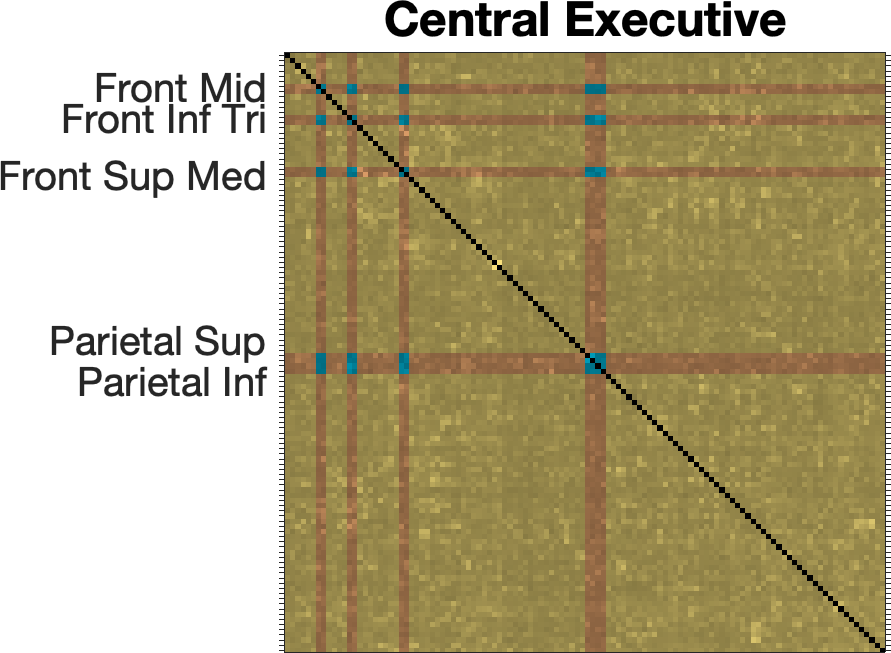}\hfil
\includegraphics[height=0.20\textwidth]{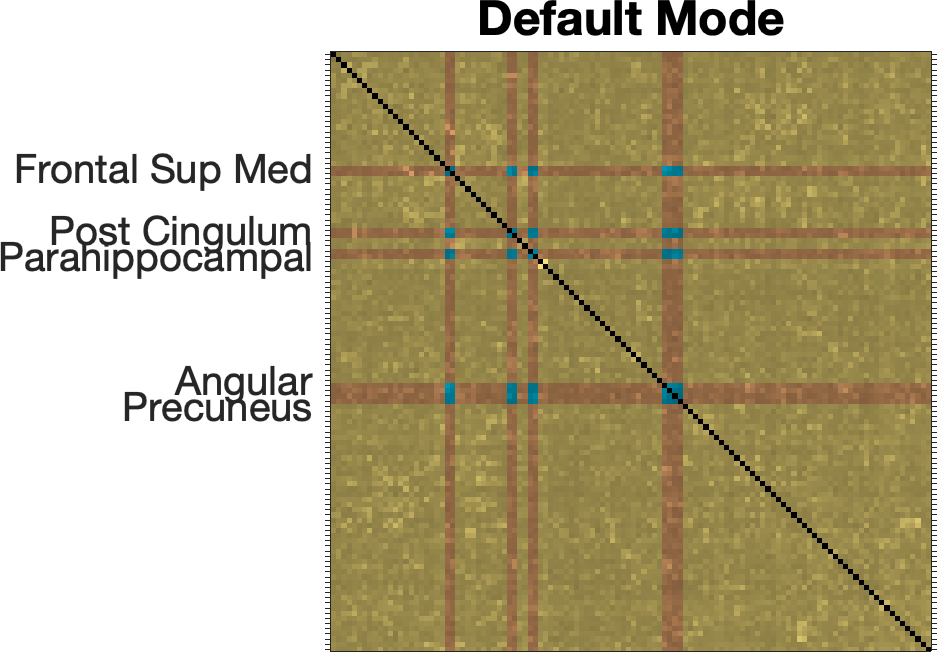}\hfil
\includegraphics[height=0.20\textwidth]{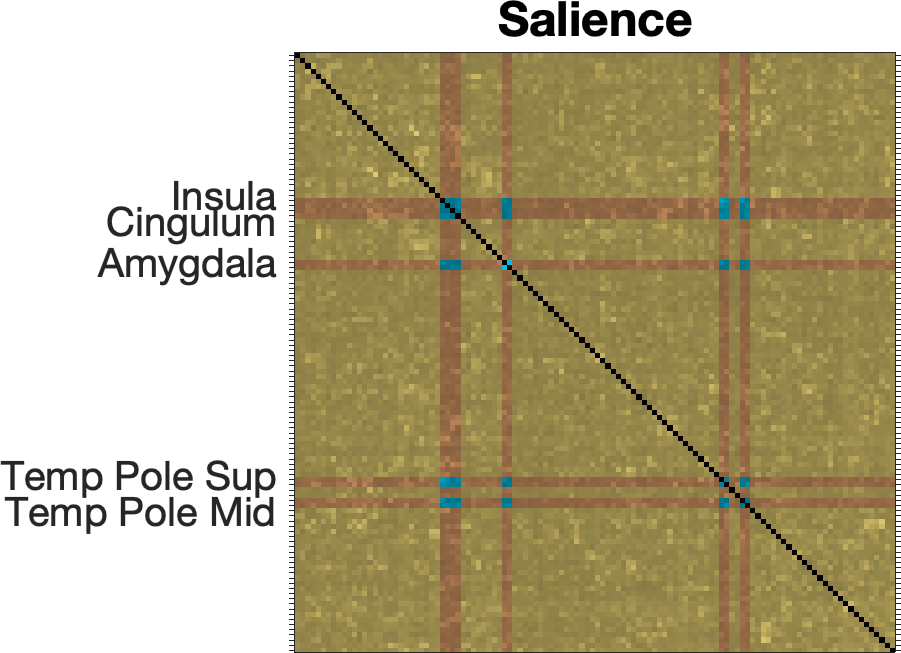}
\hfil\\
\RotText{\phantom{0000}\parbox{.32\linewidth}{\centering{\textbf{Rest}}}}
\includegraphics[width=0.3\textwidth]{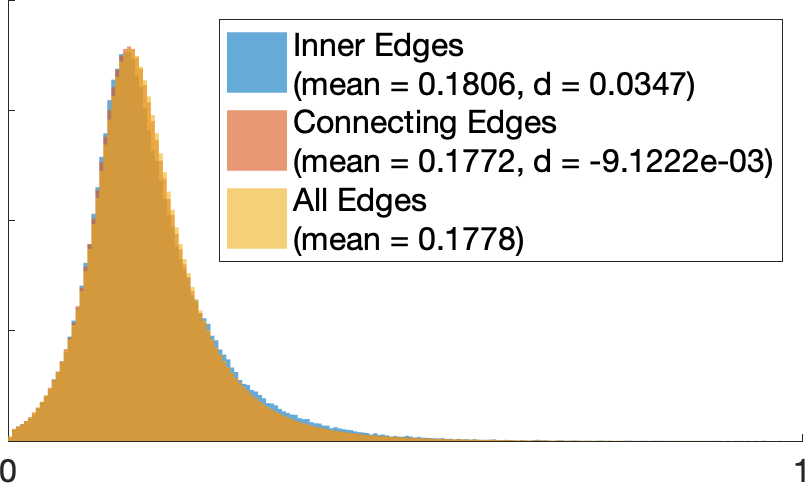}
\includegraphics[width=0.3\textwidth]{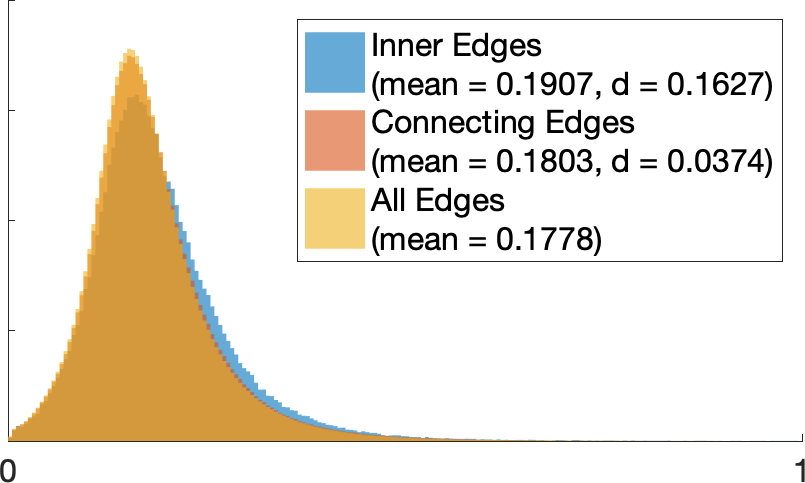}
\includegraphics[width=0.3\textwidth]{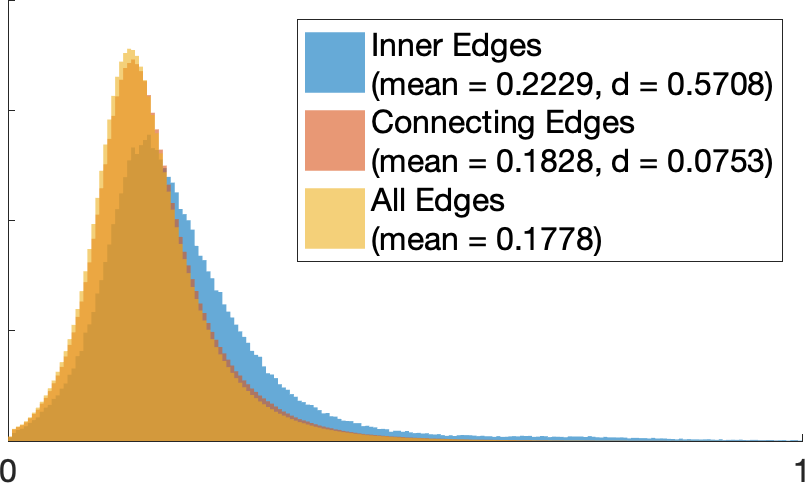}\\
\RotText{\phantom{0000}\parbox{.32\linewidth}{\centering{\textbf{Task}}}}
\includegraphics[width=0.3\textwidth]{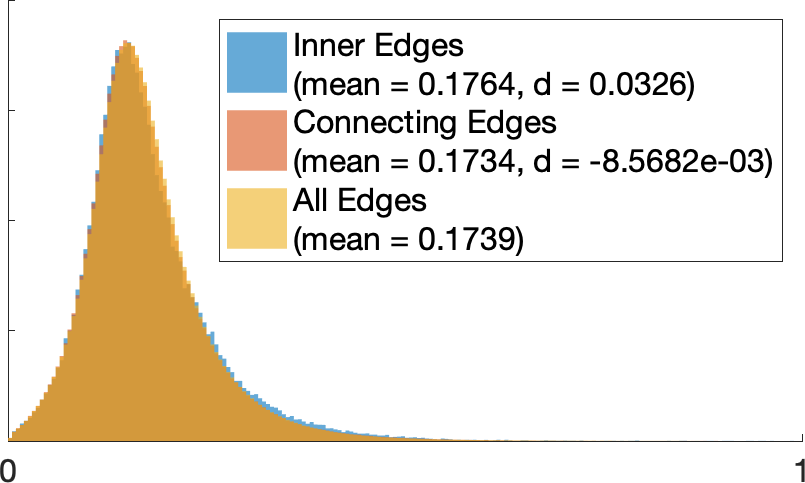}
\includegraphics[width=0.3\textwidth]{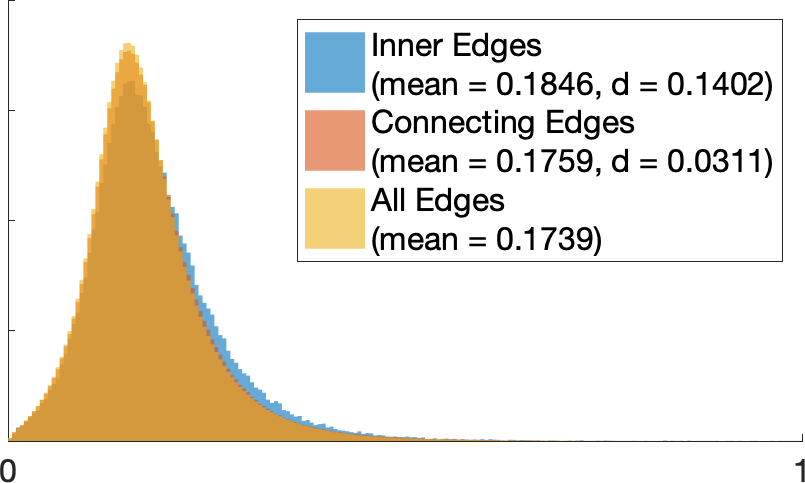}
\includegraphics[width=0.3\textwidth]{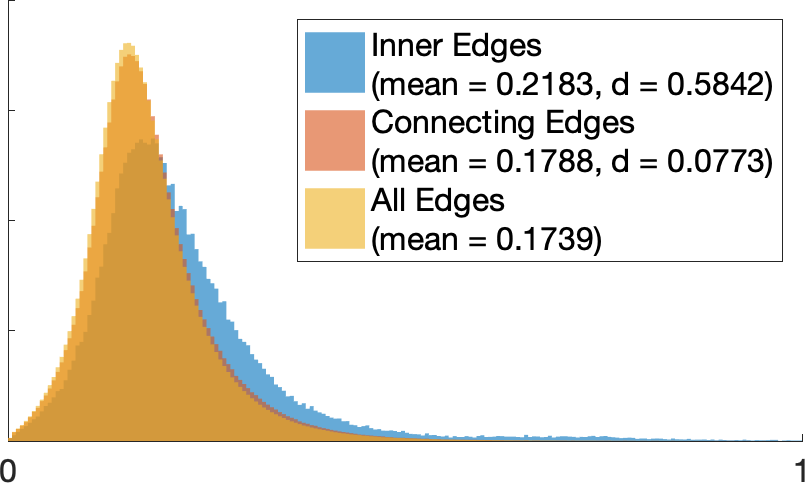}
\caption{(Top) The averaged class activation maps (CAMs) across all subjects for the complete graph classification, with the three studied networks highlighted. Area names in the AAL atlas are given. (Bottom) Histograms of all inner and connecting CAM values of the three networks, both in resting-state and task subjects, compared to the overall distribution of CAM values. Because the large number of samples, we display the effect size (measured by Cohen's d) of both inner and connecting edges compared to the CAM values of the rest of the edges.}
\label{fig:class_activation_maps_histograms}
\end{figure*}

In total, 14,683 unique connectomes (comprised of both resting-state and task data) were classified by gender across 300 ensemble models. For each connectome, a single, $116\times 116$ gradient class activation map (with $115\times 58$ unique values) was derived that indicated the general importance each particular edge played into the classification of that participant.

The distribution of edge values from CAMs, both from edges within, and edges connected to the respective networks, are shown for task and resting-state data in Figure \ref{fig:class_activation_maps_histograms}. These distributions were compared to the relative distribution of all edges with aggregated values of $115\times 58$ CAM values inside and outside of a priori networks, across 14,683 unique subjects, totalling just under 100 million values. Effect size were reported (as Cohen's d; see Figure \ref{fig:class_activation_maps_histograms}).

The differences in CAM values of edges inside and outside the CEN were non-significant, while some effects were observed for the inner, but not connecting edges of the DMN. The largest effect was seen in the salience network, having an effect size of $d > 0.57$ for task- and resting-state data separately. In CAMs overall, there were no significant differences between task- and resting-state edge values. This likely indicates that CAMs, while useful for showing which networks are important to the overall task of gender classification, are not useful for showing whether these networks were more or less important for resting-state or task data.

\subsubsection*{Occlusion}
\begin{figure*}[!t]
\centering
\includegraphics[width=0.95\textwidth]{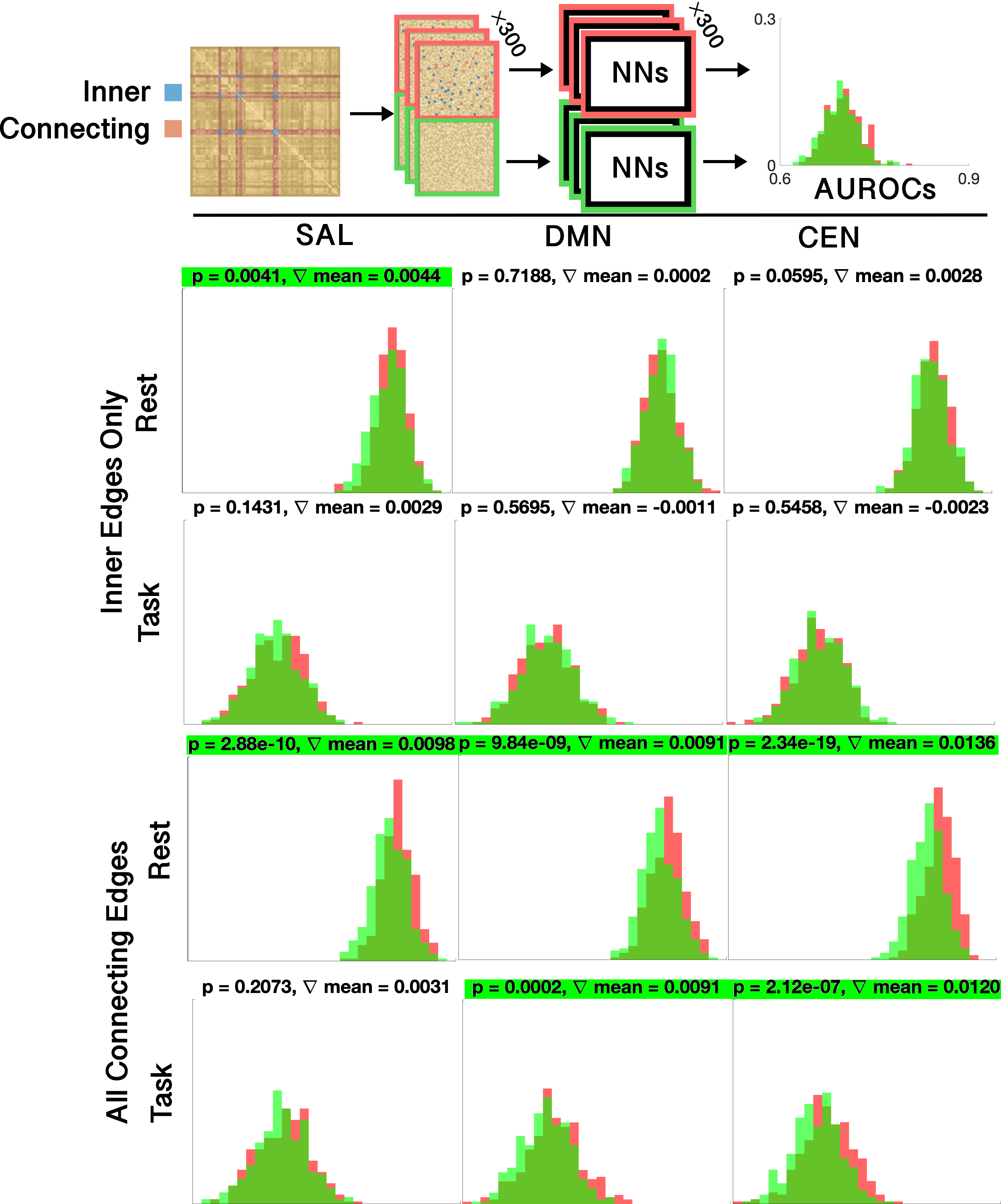}
\caption{The effects of selective network occlusion on model accuracy. (Top) the process by which occlusion AUROCs are estimated; either all inner edges of a given network, or all edges connecting to a network, are selected. The network edges are then scrambled (see Figure \ref{fig:nn_illustration}), and the selected edges are placed among one half of the scrambled edges, and in the other half left out. These two sets are then trained on $2\times300$ independent neural networks, and the resulting AUROCs are compared. (Bottom) The results. Considering only inner edges, the only statistically significant effect, after Bonferroni-Holmes correction, was the salience networks on resting-state data. Considering all connecting edges, all three networks had a significant effect on the classification of gender in resting-state data, while only the central executive network appeared to have an effect. The nonparametric Mann-Whitney U-test was used to test for statistical significance. Final model means and ensemble results are shown in Table \ref{tab:table1_ieee}.}
\label{fig:occlusion_histograms}
\end{figure*}

Using the same dataset for the gender classification task, we compared the AUROCs of 300 independent models that classified a random half of the network's edges. One set of 300 deliberately included the set of edges that constituted a network, and the other set of 300 excluded the same edges (Figure \ref{fig:occlusion_histograms}, top). By comparing the AUROCs and finding a statistically significant difference, we could assess the influence of a particular network on the classification.

The relative classification AUROCs from the halves of edges that included edges both inside and connecting to the DMN, CEN, and salience networks, as well as models completely excluding them, are shown in Table \ref{tab:table1_ieee}, while Figure \ref{fig:occlusion_histograms} shows the distribution of AUROCs on 300 models including and excluding each network, for resting-state and task data.

When considering only the edges within a network (consisting of $\frac{45}{58*115} = 0.67\%$ of total edges), modest losses in accuracy were observed (Figure \ref{fig:occlusion_histograms}), but the only one that achieved statistical significance in a Mann-Whitney U-test after Bonferroni correction was the salience network classification in resting-state data. However, when excluding all edges connected to a network (consisting of $\frac{1105}{58*115} = 16.57\%$ of total edges), a difference between resting-state and task data was observed: exclusion of all three networks led to statistically significant ($p<0.05$) decrease in AUROC for the classification of resting-state data, while the exclusion of the central executive and default mode, but not the salience networks, led to a statistically significant drop in AUROC.

\section{Discussion}

\subsection{Deep learning model}

Because it is able to capture nonlinear patterns across complex datasets, deep learning is a powerful tool for characterising biological data. However, because of interest in identifying patterns discovered by deep learning models, the interpretability of the model is just as important as performance, though it is far more difficult to quantify or even define \cite{DoshiVelez2017}. The primary methodological contribution of this study is a model that captures the contributions of individual functional connections to fMRI deep learning classification, while the results of our data show that utilisation of this model in the context of network neuroscience can shed light on between-gender differences in task- and resting-state brain networks.

Our model addresses an important problem unique to the issue of classifying graphs in CNNs, which is bias inherent in its encoding. There is no universal consensus on a method of encoding graphs for ML, though others have been proposed \cite{Jie2013,Kawahara2017,Nikolentzos2017,Kriege2019,Tixier2017,Leming2020}. Whether encoding them randomly is the optimal method for classification accuracy is up for debate, though random encoding does avoid the problem of overfitting that is present in fully-connected neural networks, and it avoids bias in the output CAMs that results from using filters with a consistent shape. In other words, the use of linear filters results in whole rows or columns of a functional connectivity matrix being emphasized, rather than particular edges. Additionally, the training scheme helped to eliminate bias from the output CAMs. Simple averaging over a large number of models and stratified cross-validation \cite{Kohavi1995} is just as important as the model architecture itself, because this allows for reduced bias from both confounding factors and natural variations in the output of nondeterministic deep learning models.

Respectively, the average AUROC for gender classification across all 300 models was 0.8010. When aggregated as an ensemble, the combined AUROC was 0.8459. This represents an improvement over our previous gender classification in \cite{Leming2020} which achieved an AUROC of 0.8295 on BioBank data (0.7683 across all datasets used) with a vertical-filter CNN balancing by only age and site. Nonetheless, due to the different balancing schemes, these two studies likely used a moderately different subset of the overall data, and so a direct comparison between the present stochastic and the previous vertical filter models in terms of accuracy is not strictly valid. Comparisons to other state-of-the-art ML studies are also not possible, since there is high variation in classification accuracy depending on how data was collected and processed \cite{Leming2020}, and few imaging studies have attempted a gender ML task on a dataset of this size.

Our training and multivariate class balancing schema, when combined, offered another uniquely important contribution. By only inputting to smaller, independent models subsets of data in which measurable confounding factors were balanced beyond any detectable statistical significance, we were able to effectively regress out any confounding factors that we were able to measure. However, by combining these subsets over a large number of independent models that were then combined in an ensemble, we were able to utilize the majority of the overall data in the end result without losing the effects of balancing. This allows us to be sure that our ML model utilized the majority of an imbalanced dataset, without achieving higher accuracy due to any confounding factors, particularly head motion and intracranial volume.

Although the balancing techniques employed prevented our model from gaining higher accuracy due to confounding factors such as age, head size, and motion, this does not necessarily mean that such differences had no influence. Class balancing does not prevent the model from internally separating data based on such factors and considering them (wholly or partially) independently. To illustrate this issue, we briefly present an analogy: consider a ML task in which pictures of different species of cat must be separated from pictures of different species of dog; such a model would likely identify generalized differences between each (e.g., the ear shape), while also containing internal representations of each type of cat and dog contained in the training set, relying on features unique to each individual species (e.g., stripes on a tiger). For instance, black fur color may be considered salient, even though it doesn't necessarily help to separate cats from dogs, because it helps the dataset to subclassify both black panthers and black Labrador retrievers.

Nonetheless, we are confident that class balancing within a cross-validation scheme reduced the influence of differences in confounding factors. We emphasize the importance of each particular step in the ML classification to achieve the output CAMs. These are: (1) random encoding, rather than encoding based on rows or columns; (2) averaging the output of many ML models, as individual outputs have a stochastic element; and (3) stratified cross-validation using balanced subsets of the data across these models.

\subsection{Neuroscientific interpretations}

Four main neuroscientific findings stand out in our results: (1) when classifying gender, the relative AUROC for resting-state data was consistently higher than that for task data by a margin of around 0.12 (Table \ref{tab:table1_ieee}); (2) the within-network edges of the salience network were considered important for characterizing resting-state data (as indicated by both occlusion and CAM results), but not task data (as indicated by occlusion results); (3) edges connecting to all three networks were important in characterizing resting-state fMRI, and notably, even when only considering edges within the networks the p-values for differences between occlusion runs were hardly above 0.05 (Figure \ref{fig:occlusion_histograms}); (4) edges connected to the CEN were the only ones that proved important to the classification of both task- and resting-state data together (Figure \ref{fig:occlusion_histograms}), even though there was little difference in the distribution of CAM values between them (Figure \ref{fig:class_activation_maps_histograms}).

The significantly lower classification accuracy of task data overall compared to resting-state data was consistent both when using complete input data and using partial input data (Table \ref{tab:table1_ieee}). The most straightforward interpretation of this result is that, in task processing, female and male brain function is more similar than it is in the resting-state. Because resting-state brain connectivity varies more than task connectivity \cite{Elton2015}, this disparity may also be due to a lower number of distinguishing features.

Explaining the apparent contradiction between our two methods regarding the status of the CEN is complex. Judging from the occlusion results, the CEN is an important network when classifying resting-state data and the only network important in classifying task data, though this is not reflected in the CAMs. Given that these two methods are established visualization methods in ML and a methodical error is unlikely, the takeaway of this contradiction is that these methods are not interchangeable and must be interpreted in their own right. The contradiction could possibly be due to a relatively small number of very salient edges connecting to the CEN, which can be seen in the right tail of the histogram in Figure \ref{fig:class_activation_maps_histograms}, though this is a very minor effect. This also shows that the interpretation of specifics in these results ought to be approached cautiously, given how novel these methods are in their application to neuroscience. Put informally, CAMs show which components of input data the deep learning model pays attention to, while occlusion shows how important a component is to the classification of a specific datapoint. With this in mind, the similar distribution of CAM values over spatially invariant task- and resting-state input data (see the histograms in Figure \ref{fig:class_activation_maps_histograms}) is not surprising since a ML model may find a particular edge salient because it might help it to internally subclassify the dataset by resting-state or task. Thus, CAMs may illustrate that a particular edge is important in the overall classification of the model, though not whether it helps in classifying a specific dataset.

With regards to the salience network, however, the two methods paint a more straightforward picture, since the inner edges of the salience network were clearly the most significant, according to CAMs (Figure \ref{fig:class_activation_maps_histograms}). Furthermore, it was the only network with inner edges that proved to be statistically significant to the classification of resting-state data (Figure \ref{fig:occlusion_histograms}). This effect may be due, in part, to the particularly salient connection between the left and right amygdala (Figure \ref{fig:class_activation_maps_histograms}) which yielded the highest CAM value by far. The difference between men and women in amygdala response has been controversial \cite{Andreano2014}, with studies disagreeing over whether there is greater activity in men \cite{Schienle2005,Goldstein2010,Sergerie2008} or women \cite{Klein2003,McClure2004,Hofer2006,Domes2010} in response to affective scenes. While CAMs cannot comment on this issue, other studies have found no difference in function at all \cite{Wrase2003,Caseras2007,Aleman2008}, which our results refute. While we can conclude from these results that the salience network is engaged differently between males and females in, at least, the resting-state, a disproportionately high value of one of its edges may drive this classification, and thus the robustness of this results requires independent verification.

The DMN is also engaged in gender differences. As can be seen from the middle histogram in Figure \ref{fig:class_activation_maps_histograms}, many of its inner edges have a higher class activation than other edges, while excluding it and all edges connected with it had a uniquely negative effect on classification (Figure \ref{fig:occlusion_histograms}). What is surprising, however, is that the DMN, which is commonly cited as the marker of resting-state functional connectivity \cite{Raichle2001} and has previously been implicated in big data gender difference studies \cite{Ritchie2018} as an area of particular interest, does not stand out from the other two networks studied. While it is not surprising that, in our occlusion tests, the CEN had a greater effect than the DMN in task classification, both tests show that, as stated above, the salience network appears to be more important and have a greater effect on classification accuracy of the resting state. This may be due to the use of a priori tests in other studies that specifically account for the DMN, the non-inclusion of subcortical areas in other studies, or the inclusion of the critical amygdala connections in the salience network, or other unknown reasons.

\section{Conclusion}

Our results show that the distinction of males and females in resting-state takes into account all of the major brain networks, particularly the salience network, which may be as a result of increased variance in resting-state networks than task-based networks, potentially offering the model a larger set of distinguishing markers. When only considering task or, more specifically, the emotional faces recognition task of the UK Biobank, areas connecting to the DMN and, more so, the CEN showed significantly altered function, while function of the the salience network was not different enough to significantly aid in single-subject classification (Figure \ref{fig:occlusion_histograms}). Methodologically, we have also shown the applicability and limitations of two different ML visualization methods to brain network data, as well as ML's applicability to big data in a scientific field.


%



\section{Acknowledgment}

This research was co-funded by the NIHR Cambridge Biomedical Research Centre and Marmaduke Sheild. Matthew Leming is supported by a Gates Cambridge Scholarship from the University of Cambridge. Lena Dorfschmidt provided advice on several aspects of gender differences in the brain for this study.

This research has been conducted using the UK Biobank Resource [project ID 20904], co-funded by the NIHR Cambridge Biomedical Research Centre and a Marmaduke Sheild grant to Richard A.I Bethlehem and Varun Warrier. The views expressed are those of the author(s) and not necessarily those of the NHS, the NIHR or the Department of Health and Social Care.

\ifCLASSOPTIONcaptionsoff
  \newpage
\fi

\bibliographystyle{IEEEtran}
\bibliography{IEEEabrv,references}

\end{document}